\documentclass[preprint2]{aastex}

\newcommand{\axaf}{\mbox{\em Chandra\/}}
\usepackage{graphicx}
\usepackage{fancyheadings}
\usepackage{color}

\shorttitle{X-ray Spectrum of SDSS1306}
\shortauthors{Schwartz and Virani}

\begin{document}

\title{\emph{Chandra} measurement of the X-ray spectrum of a quasar at z=5.99}
\author{D.~A. Schwartz, S.~N. Virani}
\affil{Harvard-Smithsonian Center for Astrophysics, Cambridge, MA 02138}

\email{das@head.cfa.harvard.edu}

\begin{abstract}

We report the first measurement of the X-ray spectrum of the z
= 5.99 quasar SDSSp~J130608.26+035626.3 from a 120 ks observation by
the \axaf\ ACIS-S instrument.  Between 0.5 and 7 keV, corresponding to
3.5--49 keV in the quasar rest frame, we find an energy index of 0.86
$\pm$ 0.2, consistent with the typical indices found for radio quiet
quasars at lower redshifts, and inconsistent with the index required
to match the diffuse X-ray background.  We have a weak  indication of a
redshifted Fe-K line.  In comparing the
counting rate between an earlier, short observation and the longer
observation reported here, we find evidence for source variability at
the 99.9\% confidence level. We note that other nearby X-ray sources
would bias the measured $\alpha_{\rm ox}$ = 1.70 by $-$0.09 if the
X-ray flux were determined from within a 60\arcsec\ extraction
circle. Our results for the energy index and the $\alpha_{\rm ox}$ are
consistent with no strong evolution in the active galactic nucleus emission mechanism with
redshift out to z $\approx$ 6, and therefore with the picture that
massive black holes have already formed less than 1 Gyr after the big
bang.

\end{abstract}

\keywords{quasars: general--- quasars: individual (SDSSp J130608.26+035626.3)--- X-rays: galaxies}

\section{INTRODUCTION}

The Sloan Digital Sky Survey has revealed the most distant quasars
known \citep{Fan01, Fan04}.  \axaf\ observations on 2002 January 29
easily detected three quasars at redshifts 5.82, 6.28, and 5.99 in
brief, $\approx$ 5--8 ks, ACIS-S observations \citep{Schwartz02,
Brandt02, Wilkes02, Bechtold02}. The latter, SDSSp J130608.26+035626.3
(hereafter, J1306), was followed up with a 120 ks \axaf\ observation,
based on the claim by \citet{Schwartz02} of an associated
jet. \citet{Ivanov02} subsequently identified the X-rays attributed to
a jet with a faint galaxy, and the present observation clearly
confirms the galaxy identification.

From the optical observations, J1306 is a broad line quasar, whose
spectrum contains Ly$\alpha$+NV emission, and various intervening
absorption systems \citep{Fan01, Becker01}. This is a general
characteristic of the 12 quasars now known at z $\approx$ 6
\citep{Fan04}. From the optical absolute magnitude M$_{1450}$ = $-$26.93
\citep{Fan01}, the central black hole mass must be at least 10$^{{\rm
9}}$ M$_{\sun}$, unless it is emitting at much greater than the Eddington
luminosity.

We report here a study of the X-ray variability and spectrum of
J1306. The X-ray spectrum of distant quasars is of interest to search 
for possible evolution of the active galactic nucleus (AGN) emission mechanism, which might give
clues to the history of the formation and the fueling of the central
massive black hole.  With  AGNs widely inferred to make
up the bulk of the diffuse X-ray background, it remains to define the
model of quasar evolution and of the relative numbers of absorbed and
unabsorbed emitters that will reproduce the shape of the 2--10 keV
diffuse X-ray background to the precision measured by
\citet{Marshall80}. It remains possible that a high fidelity spectral
fit must also consider a contribution from quasars with concave
curvature in their spectrum from 5 to 40 keV, as suggested by
\citet{Schwartz88}. 

Previous measurements of radio-quiet quasars in the redshift range
2.64--4.35 gave photon spectral indices $\Gamma \approx$ 2.0
\citep{Ferrero03,Brocksopp04,Grupe04}, with the exception of an index
1.4 reported by \citet{Brocksopp04} for RX J122135.6+280613 at
z=3.305. \citet{Yuan98} report a mean index $\Gamma \approx \rm{
2.23\pm 0.15}$ for 56 \emph{ROSAT} quasars with z $>$
2.0. The result reported here provides an example at z=5.99 with
$\Gamma$= 1.86. 

\section{OBSERVATIONS OF J1306}
\label{sec:obs}

J1306 was observed with the back-illuminated S3 CCD of the Advanced
CCD Imaging Spectrometer (ACIS) at the nominal aim point for
approximately 118 ks live time starting on 2003 November 29. The
\axaf\ data were reduced in the standard manner using CIAO version 3.0.2 and
the latest calibration database (CALDB) release (ver. 2.26; released 2004
February 2) which incorporates an improved ACIS contamination
calibration file and corrects a one-channel offset in the ACIS S3
response matrix.  The contamination correction is required since the
degradation of the effective ACIS throughput is strongly energy
dependent. Only events for \emph{ASCA} grades 0, 2-4, and 6 were retained for
analysis. We restrict the energy range to 0.5--7.0 keV, as the
background rises steeply below and above those limits.\footnote{See:
http://cxc.harvard.edu/contrib/maxim/stowed/} However, no photons
greater than 5.1 keV were detected. We applied
``corr\_tgain.1.0''\footnote{See the CXC Contributed Software Exchange
Web page at http://asc.harvard.edu/cgi-gen/cont-soft/soft-list.cgi} to
the event list to correct the measured X-ray energy for the secular gain
drift which is primarily caused by gradual deterioration of the charge
transfer inefficiency of the ACIS CCDs.  We checked that no background
flares occurred during the observation. We also inspected the
individual photon arrival times from the quasar and there is no
evidence for transient phenomena affecting the data.

\begin{figure}[h]
\begin{minipage}[c]{0.44\textwidth}
\includegraphics*[width=\textwidth]{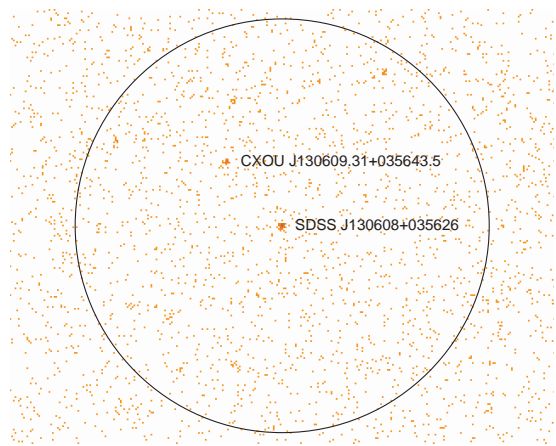}
\caption{\label{fig:field} Circle indicates a 1\arcmin ~radius around
J1306. The pointlike source CXOU~130609.31+035643.5 is the faint
\emph{I}-band galaxy found by \protect\cite{Ivanov02}. Several other weak sources
appear in the region, and they would bias the measured $\alpha_{ox}$ in
an observation with much poorer angular resolution.}
\end{minipage}
\end{figure}

\begin{figure}[h]
\begin{minipage}[c]{0.44\textwidth}
\includegraphics*[width=\textwidth]{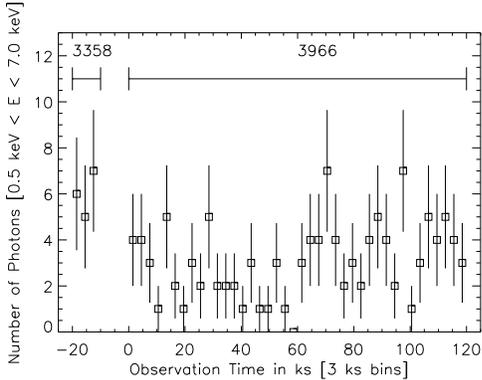}
\caption{\label{fig:lightcurve} Light curve of J1306 during the
  short observation (ObsID 3358) and the present observation (ObsID
  3966). The gap between the two observations is 22 months. We plot the
  total number of photons between 0.5 and 7 keV, within a 3\arcsec\
  radius of the quasar, sorted into 3 ks bins.
}
\end{minipage}
\end{figure}

Figure~\ref{fig:field} shows the field within 1\arcmin ~of the
quasar. CXOU~130609.31+035643.5, a point source approximately
24\arcsec ~to the northeast from J1306, is the faint \emph{I}-band galaxy found by
\cite{Ivanov02} and now confirmed as the correct identification rather
than the suggestion of a jet \citep{Schwartz02}.
Figure~\ref{fig:lightcurve} shows the light curve for J1306 during the
earlier short observation (ObsID 3358) and the present long
observation (ObsID 3966). We detected 18 photons between 0.5 and 7.0
keV in 8.16 ks during the 2002 January observation. However, based on
the counting rate of the present observation, only 8 photons are
expected. The probability of observing 18 photons when 8 are expected
is only 0.1\%. Therefore, this suggests that we see intrinsic
variability in the source between these two epochs.  Weak evidence of
variability within the current observation would require better
statistics to confirm. Note that the elapsed time of 22 months between
the two observations corresponds to a time scale of 3.15 months in
the source frame.


\section{THE X-RAY SPECTRUM}

The X-ray spectrum was extracted using a circular region with a
3\arcsec ~radius centered on the brightest pixel.\footnote{These data
and appropriate background were made publicly available immediately
following the observations at
http://asc.harvard.edu/acis/dschwartz.html}   The background was taken from an annular region centered on
the quasar (inner radius of 28\arcsec ~and outer radius of 80\arcsec)
that is free of obvious sources. The number of net source counts in
the 0.5--7.0 keV band after background subtraction is 118 photons (122
photons detected in total and used for the spectral analysis), giving
a net count rate of $(9.97 \times 10^{-4}) \pm (9.72 \times 10^{-5})$
$\rm{counts~ s^{-1}}$.  This is consistent with a point response,
where we find 114 X-rays  within a 2\farcs5 diameter circle, which will enclose 95\% of
photons at 1.5 keV \citep{Jerius00}, an energy near the median of our
spectrum.

\begin{figure}[h]
\includegraphics*[width=.3\textwidth, angle=-90]{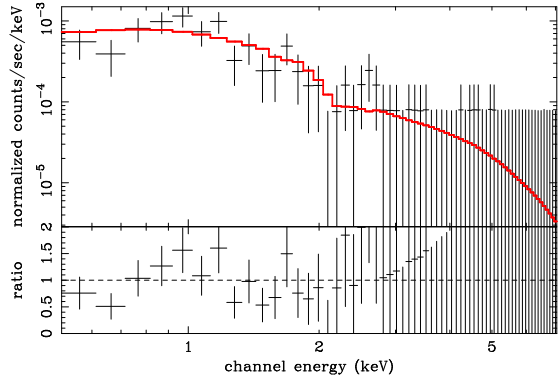}
\caption{\label{fig:pow} X-ray spectrum of J1306 observed between 0.5 and
  7.0 keV. The solid histogram gives the best fit power law model, with
  absorption fixed at the Galactic value.}
\end{figure}

The photons were grouped, using the CIAO tool \emph{dmgroup}, into 100 eV
bins. Because of the small number of photons in most bins, the Cash
statistic \citep{Cash79} was used as the figure of merit in spectral
fits performed within XSPEC \citep{Arnaud96}.  We fit the spectrum to
a power law, with galactic absorption fixed at its measured value of
$2 \times 10^{20}$ H-atoms cm$^{-2}$ \citep{Stark92}. This is shown in
Figure~\ref{fig:pow}. The fit is obviously adequate, and gives an
energy spectral index $\alpha =$ 0.96 $\pm$ 0.18. The Galactic
absorption-corrected flux in the observed 0.5--7.0 keV band is $6.1
\times 10^{-15}$ $\rm{ergs ~cm^{-2} ~s^{-1}}$.  For $H_{0}$ = 71 km
~s$^{-1}$ ~Mpc$^{-1}$, $\Omega_{m}$ = 0.27, and $\Omega_{\Lambda}$ =
0.73, we derive an X-ray luminosity of 2.2$ \times 10^{45}$
$\rm{ergs ~s^{-1}}$ emitted in the source frame 2--10.0 keV band.

\begin{figure}[h]
\includegraphics*[width=.3\textwidth,
  angle=-90]{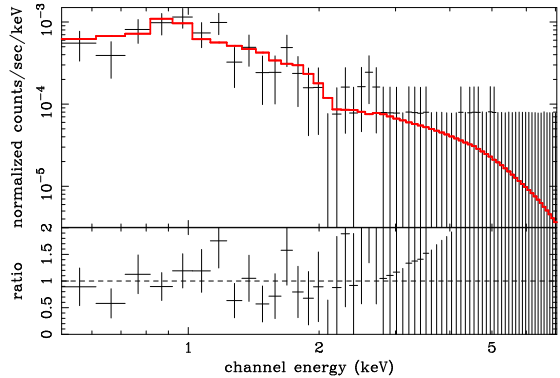} 
\caption{\label{fig:line} Same as Figure~\ref{fig:pow} but with a
 redshifted 6.4 keV Fe-K 
 emission line.  The line energy is fixed at 0.916 keV, and the
 intrinsic width of the 
 Gaussian is fixed to 0.}
\end{figure}

\begin{figure}[h]
\includegraphics*[width=.35\textwidth, angle=-90]{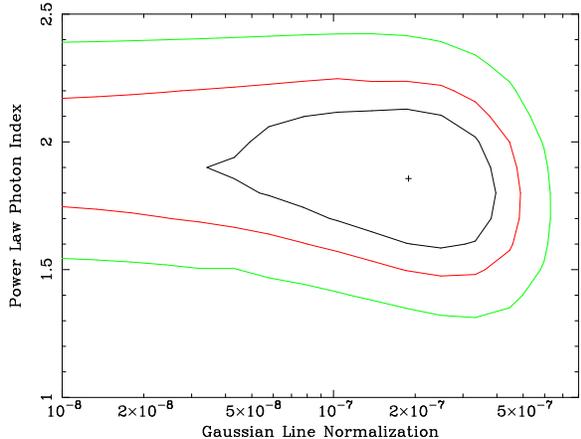}
\caption{\label{fig:contours}Contours give the joint 68\%, 90\%, and
  99\% confidence for the two interesting variables, photon index $\Gamma_X$
  and the amplitude of the Fe-K line. For the line normalization in
  units of photons cm$^{-2}$ s$^{-1}$, 10$^{-7}$ corresponds to an
  equivalent width of 60 eV.}
\end{figure}

The residuals apparent below 1 keV hint at the possibility of
redshifted Fe-K emission.  If we add a narrow Gaussian line to the fit
model, with the line energy fixed at 0.916 keV for redshifted neutral
iron and the line $\sigma$ fixed at the intrinsic S3 resolution at the
source position on the CCD, we derive the fit shown in
Figure~\ref{fig:line}. The power-law energy index is slightly flatter,
$\alpha$=0.86 $\pm$ 0.20, and the power-law normalization decreases by
about 10\%.  The Cash statistic decreases by about 3.3. Although this
would formally be 93\% significance \citep{Cash79},
\citet{Protassov02} point out that the likelihood ratio test is not
valid for such an additive component to a spectral model. The chance
that Poisson statistics leads to the residuals between 800 and 1000 eV
in Figure~\ref{fig:pow} is 12.7\%, so clearly further confirmation is
needed for this feature. If real, the equivalent width of the the Fe-K
emission line would be approximately 114 eV.
Figure~\ref{fig:contours} shows the joint contours for the X-ray slope
$\alpha_X$ and the amplitude of the Fe-K line.

We compute $\alpha_{\mathrm{ox}}$ according to the original definition of
\citet{Tananbaum79},
\begin{equation}
\label{eq:alfox}
 \alpha_{\mathrm{ox}} =-\log(S_{x}/S_{o})/\log(\nu_{x}/\nu_{o}), 
\end{equation}
where $S_{x}$ is the X-ray
flux density at 2 keV, $\nu_{x}$= 4.8$\times$10$^{17}$ Hz, and $S_{o}$
is the optical flux density at 2500 \AA,
$\nu_{o}$=1.2$\times$10$^{15}$ Hz. We calculated the optical
monochromatic luminosity at rest-frame 2500 \AA ~using the observed
optical flux \citep{Fan01} and assuming a power law optical continuum 
slope with $\alpha_{\mathrm{o}}$ = 0.5. This yields a value for
$\alpha_{\mathrm{ox}}$ of 1.70. This
result is consistent with the values determined in the earlier
 observations, with small differences due to the fact that the
 previous conversion of counts to flux density  merely assumed an
 energy index of 
 0.7 or 1.0 by \citet{Schwartz02} and \citet{Brandt02}, respectively,
 and that the source flux in the present observation is
 only half that in 2002 January.
We note that an X-ray flux extraction circle of 
60$\arcsec$ radius would lead to a decrease in
$\alpha_{\mathrm{ox}}$ to 1.61, because of the inclusion of flux from other
nearby X-ray sources, principally CXOU~130609.31+035643.5, as can be
seen in Figure~\ref{fig:field}. 


\section{DISCUSSION}

In this Letter, we present the first \axaf\ X-ray spectrum of a z
$\sim$ 6 quasar.  The values for the energy index and the
$\alpha_{\mathrm{ox}}$ derived from this \axaf\ spectrum of J1306 are
consistent with the values found for lower redshift radio-quiet
quasars \citep{Brandt02}. Specifically,  our result for
$\alpha_{\mathrm{ox}}$ is consistent with no strong evolution in
$\alpha_{\mathrm{ox}}$ for optically selected radio-quiet quasars out
to z $\approx$ 6.  A preprint by \citet{Farrah04} gives an \emph{XMM}
spectrum of SDSS J1030+0524, a quasar at z=6.30. They reach very
similar conclusions as to the spectrum and luminosity, and
implications for lack of evolution.  While \emph{XMM} affords higher
statistics, the 45\arcsec ~extraction radius gives less certainty on
the precise origin of the flux. Thus the two observations are somewhat
complementary.  

X-ray studies of high-redshift quasars reveal the conditions in the
immediate vicinity of their supermassive black holes. Our spectrum of
J1306 hints at the presence of an Fe-K emission line. Our poorly
determined EW of 114 eV is consistent with the scatter in compilations
by \citet{Nandra97} and \citet{Page04} which would extrapolate to
predict a $\approx$70 eV EW for a radio-quiet quasar of the
2$\times$10$^{45}$ ergs
s$^{-1}$ luminosity. Recent near-infrared spectroscopy of three other high
redshift SDSS quasars (z $\geq$ 5.7) with the \emph{Hubble Space Telescope} \citep{Freudling03}
clearly detected Fe II  emission lines at 2200-2600 \AA ~(rest frame)
and the Keck-II observations of \citet{Barth03} have detected Si IV, C
IV, Mg II, and the 2900-3000 \AA ~Fe line complex. Both results
indicate that the strength of the line complex, as well as the Fe
II/Mg II ratio, is comparable to that of lower redshift quasars. The
results of near-infrared studies such as these as well as the results of X-ray
studies such as the one presented here and elsewhere \citep{Brandt02,
Vignali03} are beginning to indicate that there apparently is no
strong evolution in the broadband AGN emission mechanism with redshift
out to z $\approx$ 6.  This conclusion, if confirmed with a larger
sample, implies that formation of the central black hole engines can
be completed in less than 1 Gyr after the big bang, with the inference
that a massive galaxy in which they are imbedded has also
formed. Therefore, in order to address the question of which, if any,
of the observed quasar properties evolve with time, we must probe
deeper still to identify quasars at even higher redshifts.  This might
best be done with the \axaf\ deep surveys, e.g., by identifying the
objects with extreme X-ray-to-optical flux ratios found by
\citet{Koekemoer04}.

\acknowledgments This work was supported in part by NASA contract
NAS8-39073 to the \emph{Chandra} X-Ray Center (CXC)  and CXC grant
G03-4120X  to the Smithsonian Astrophysical Observatory.

\end{document}